\documentclass[twocolumn,english,aps, reprint, twocolumn]{revtex4-1}
\usepackage[LGR,T1]{fontenc}
\usepackage[latin9]{inputenc}
\usepackage{color}
\usepackage{textcomp}
\usepackage{graphicx}
\usepackage{setspace}
\usepackage{subscript}

\makeatletter

\DeclareRobustCommand{\greektext}{%
  \fontencoding{LGR}\selectfont\def\encodingdefault{LGR}}
\DeclareRobustCommand{\textgreek}[1]{\leavevmode{\greektext #1}}
\DeclareFontEncoding{LGR}{}{}
\DeclareTextSymbol{\~}{LGR}{126}

\makeatother

\usepackage{babel}
\begin{document}

\title{Collapse of the low temperature insulating state in Cr-doped V\textsubscript{2}O\textsubscript{3}
thin films }

\author{Pía Homm\textsuperscript{1}, Leander Dillemans\textsuperscript{1},
Mariela Menghini\textsuperscript{1}, Bart Van Bilzen\textsuperscript{1},
Petar Bakalov\textsuperscript{1}, Chen-Yi Su\textsuperscript{1},
Ruben Lieten\textsuperscript{1}, Michel Houssa\textsuperscript{1},
Davoud Nasr Esfahani\textsuperscript{2}, Lucian Covaci\textsuperscript{2},
Francois Peeters\textsuperscript{2}, Jin Won Seo\textsuperscript{3},
and Jean-Pierre Locquet\textsuperscript{1}}

\affiliation{\textsuperscript{1}Department of Physics and Astronomy, KU Leuven,
Celestijnenlaan 200D, 3001 Leuven, Belgium, \textsuperscript{2}Department
of Physics, University of Antwerp, Groenenborgerlaan 171, B-2020 Antwerp,
Belgium, \textsuperscript{3}Department of Materials Engineering,
KU Leuven, Kasteelpark Arenberg 44, 3001 Leuven, Belgium }

\email{Pia.Homm@fys.kuleuven.be}

\selectlanguage{english}%
\begin{abstract}
We have grown epitaxial Cr-doped V\textsubscript{2}O\textsubscript{3}
thin films with Cr concentrations between 0 and 20\% on (0001)-Al\textsubscript{2}O\textsubscript{3 }by
oxygen-assisted molecular beam epitaxy. For the highly doped samples
(> 3\%), a regular and monotonous increase of the resistance with
decreasing temperature is measured. Strikingly, in the low doping
samples (between 1\% and 3\%), a collapse of the insulating state
is observed with a reduction of the low temperature resistivity by
up to 5 orders of magnitude. A vacuum annealing at high temperature
of the films recovers the low temperature insulating state for doping
levels below 3\% and increases the room temperature resistivity towards
the values of Cr-doped V\textsubscript{2}O\textsubscript{3} single
crystals. It is well-know that oxygen excess stabilizes a metallic
state in V\textsubscript{2}O\textsubscript{3} single crystals. Hence,
we propose that Cr doping promotes oxygen excess in our films during
deposition, leading to the collapse of the low temperature (LT) insulating
state at low Cr concentrations. These results suggest that slightly
Cr-doped V\textsubscript{2}O\textsubscript{3} films can be interesting
candidates for field effect devices.
\end{abstract}
\maketitle
The metal-insulator transition (MIT) in vanadium oxides forms a topic
of intense research since many years. Not only are there many different
structural phases present in the V-O phase diagram \cite{Wriedt 1989}
but also a large distribution of MIT temperatures is observed \cite{Magneli T-MIT 2004}.
Most well-known phases are vanadium sesquioxide (V\textsubscript{2}O\textsubscript{3})
and vanadium dioxide (VO\textsubscript{2}) with transition temperatures
of 160 K and 340 K and a change in resistivity across the MIT of about
seven and four orders of magnitude, respectively \cite{McWhan PRB 1973,Morin PRL 1959}.
Because of this large change in resistivity, there has been considerable
interest to drive this transition with additional stimuli besides
the change in temperature \cite{Yang Review 2011}. Most relevant
would be to induce the transition through the application of an electric
field as in a field-effect transistor (FET) device. Electric field
induced resistive switching has been observed many times but so far
this has been mostly limited to either: i) only a small change of
the maximal resistivity, ii) local temperature changes induced by
Joule heating, iii) changes in the oxygen content and iv) the formation
of locally more conducting paths \cite{Guenon EPL 2013,Brockman Nature 2014,Ko APL 2008,Jeong Sci 2013,Madan ACS Nano 2015}.
Essentially, it has turned out that the correlated electron states
in these oxides are rather stable and can not easily be changed through
the application of an electric field.

Our ultimate research goal is to find a method to tune the properties
of vanadium oxide based compounds in order to facilitate the electric
field induced MIT. One of these methods is to change the dimensionality
towards 2D systems by using epitaxially grown thin films (TF). In
that case the lattice parameters of the film tend to change and adapt
to those of the single crystalline substrates with the electrical
properties being modified as demonstrated in TF grown with different
methods \cite{Schuler TSF 1997,Luo APL 2004,Yonezawa SSC 2004,Autier PRB 2006,Dillemans TSF 2012,Dillemans APL 2014}.
One other well-known method is to dope the oxide with different elements
such as Cr and Ti. For the case of Cr doping Frenkel \textit{et al}.
\cite{Frenkel PRL 2006} have shown evidence in single crystals (SC)
that Cr atoms create substitutional strain defects in the V\textsubscript{2}O\textsubscript{3}
lattice leading to a disordered system of bonds around the average
trigonal lattice determined by X-ray diffraction (XRD). The long range
strain field around Cr atoms results in insulating regions even at
the low Cr concentration of 1\%. In particular, for Cr concentrations
below 1.8\% besides the low temperature MIT, a paramagnetic metal
(PM) to paramagnetic insulator (PI) transition around room temperature
has been reported in bulk \cite{McWhan PRB 1973}. When increasing
dopant concentration a transition from an antiferromagnet insulator
(AFI) to a PI takes place at low temperatures. In this case, both
the low temperature resistivity (LTR) and the room temperature resistivity
(RTR) increase considerably as observed in single crystal \cite{Jayaraman PRB 1970,McWhan PRB 1970,Kuwamoto PRB 1980}
and thin film alloys \cite{Metcalf TSF 2007}. Another effective way
to manipulate the MIT is the non-stoichiometry, which in the literature
has been formulated as V\textsubscript{2}O\textsubscript{3+\textgreek{d}}
or V\textsubscript{2-y}O\textsubscript{3}, since metal vacancies
can be generated in the cation sublattice due to the filling of all
the available oxygen sites. For that case, it is well-known that the
electrical properties change drastically with increased oxygen doping
and that beyond \textgreek{d} = 0.03 the LT MIT is suppressed \cite{McWhan JPC 1971,Ueda JSSC 1980,Shivashankar PRB 1983}.
This stabilization of the metallic phase at all temperatures has also
been observed with Ti doping and application of hydrostatic pressure
\cite{Shivashankar PRB 1983} and in Cr-doped nonstoichiometric V\textsubscript{2}O\textsubscript{3}
SC as reported in \cite{Kuwamoto-Honig JSSC 1980} for 1\% Cr concentration
and around 0.04 oxygen excess.

In this work, we report the epitaxial growth of Cr-doped V\textsubscript{2}O\textsubscript{3 }TF
on Al\textsubscript{2}O\textsubscript{3 }substrates using molecular
beam epitaxy and we compare their structural properties (\textit{a}
and \textit{c} lattice parameters) with those of the bulk compounds.
Next, the electrical properties are reported with the striking result
that the low Cr-doped films show a nearly metallic behavior at all
temperatures. To elucidate the main reason for this observation, vacuum
annealing at high temperature is performed on the films, which results
in a recovery of the LT insulating state for the low Cr-doped cases. 

The Cr-doped V\textsubscript{2}O\textsubscript{3 }TF (60 to 80 nm)
have been deposited by oxygen-assisted molecular beam epitaxy (MBE)
in a vacuum chamber (Riber) with a base pressure of 10\textsuperscript{-9}mbar.
Substrates of (0001)-Al\textsubscript{2}O\textsubscript{3} were
used without prior cleaning and were slowly heated to the growth temperature
of 650\textsuperscript{o}C as measured with a thermocouple. Alloys
with Cr concentrations between 0 and 20\% have been grown by co-deposition
of V and Cr metals in an O\textsubscript{2} partial pressure of 8.2
\textminus{} 8.5 \texttimes{} 10\textsuperscript{\textminus 6} Torr,
which constitutes most of the total pressure in the chamber and is
at least two orders of magnitude higher than residual gases like H\textsubscript{2}.
V was evaporated from an electron gun with a deposition rate of 0.1
Å/s calibrated with a quartz crystal microbalance (QCM) prior deposition
while Cr was evaporated from a Knudsen cell (Veeco) by using different
Cr fluxes to obtain the particular Cr/V ratios. Very low deposition
rates were achieved by extrapolating calibration rate curves fitted
with an exponential temperature-dependence. The growth time is 60
minutes for all samples. During growth, the metal layers will combine
with the O\textsubscript{2} producing the oxide layers. This will
translate in a final oxide layer thicker than expected from considering
only the V + Cr deposition rates. As a result, the thickness of the
samples increases more than 20\% for the highest Cr concentration.
\textit{In situ} reflection high energy electron diffraction (RHEED)
is used qualitatively to confirm the epitaxy. After deposition, the
samples were characterized by means of high resolution XRD, X-ray
reflectivity (XRR) and X-ray reciprocal space mapping (RSM) using
a Panalytical X\textquoteright pert Pro diffractometer. Temperature
dependent resistivity measurements were assessed in the Van der Pauw
(VDP) configuration with Au/Cr contacts and using an Oxford Optistat
CF2-V cryostat with a sweep rate of 1.5 K per minute. After their
initial electrical characterization, the TF were annealed in vacuum
for 5 minutes at the same temperature as the deposition (650\textsuperscript{o}C).
RSM was performed after the annealing to confirm that the structural
quality of the films is preserved. Finally, the transport properties
were again measured under the same conditions as the as-grown films. 

\begin{figure}
\begin{singlespace}
\includegraphics[bb=0bp 0bp 703bp 496bp,scale=0.35]{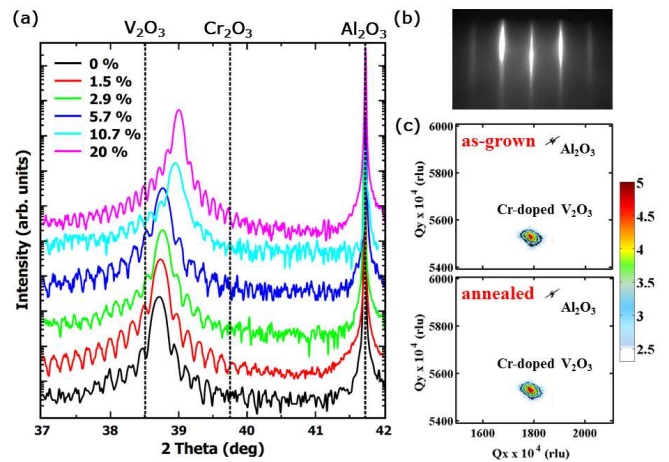}\protect\caption{\label{=0003B8/2=0003B8-scans}(a) XRD \textgreek{j}/2\textgreek{j}
scans around the ($0$$0$$0$$6$) reflection of Cr-doped V\protect\textsubscript{2}O\protect\textsubscript{3}
TF showing Pendellösung fringes. (b) RHEED image taken after deposition
along the {[}$1$$0$$\bar{1}0${]} direction of the undoped sample.
(c) RSMs of the ($1$ $0$ $\bar{1}$ $\underline{10}$) substrate
and layer peaks for the 1.5\% Cr-doped sample before and after annealing.
Qx and Qy are the components of the scattering vector along in- and
out-of-plane directions, respectively. The intensity scale is logarithmic.}
\end{singlespace}
\end{figure}

In order to examine the crystalline quality, the as-grown films were
first characterized by high resolution XRD. Figure \ref{=0003B8/2=0003B8-scans}
(a) shows \textgreek{j}/2\textgreek{j} scans in logarithmic scale
around the symmetric (0006) reflection of the Cr-doped V\textsubscript{2}O\textsubscript{3}
TF and the Al\textsubscript{2}O\textsubscript{3} substrate. Finite
size oscillations (Pendellösung fringes) around the layer peak indicate
that the films have a smooth surface and interface. The layer peak
position shifts from the V\textsubscript{2}O\textsubscript{3} to
the Cr\textsubscript{2}O\textsubscript{3} bulk \textit{c} lattice
parameter when increasing the Cr concentration, confirming the substitutional
doping of Cr in the lattice, demonstrated in SC \cite{Frenkel PRL 2006}.
The absence of extra diffraction peaks indicates that there are no
Cr\textsubscript{2}O\textsubscript{3} impurities. A RHEED pattern
of the undoped sample taken after deposition (Figure \ref{=0003B8/2=0003B8-scans}
(b)) presents clear streaks with no indications of poly-crystallinity.
RSMs around ($1$ $0$ $\bar{1}$ $\underline{10}$) the reflection
for the 1.5\% Cr-doped sample in Figure \ref{=0003B8/2=0003B8-scans}
(c), before and after the annealing, evidence that the films are single
phase and that the crystalline quality is preserved after the thermal
treatment.

\begin{figure}
\begin{singlespace}
\includegraphics[scale=0.6]{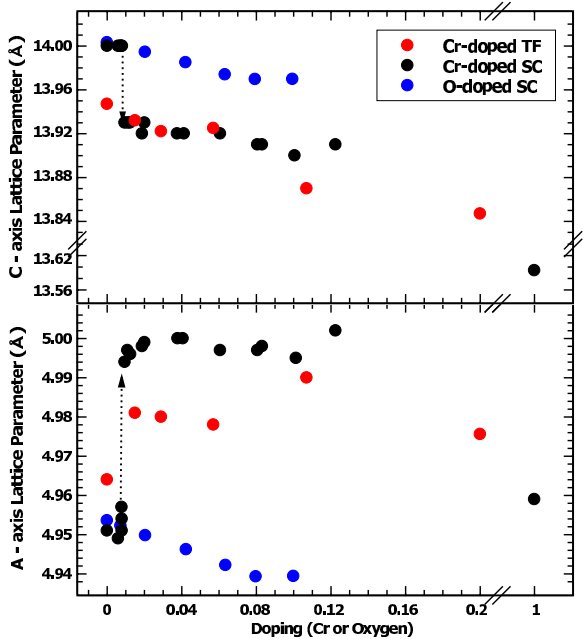}\protect\caption{\label{Lattice-SC-TF}Lattice parameters as a function of Cr doping
of thin films (TF) compared with single crystals (SC). Cr and O-doped
SC data extracted from \cite{McWhan PRB 1970} and \cite{Ueda JSSC 1980},
respectively. }
\end{singlespace}
\end{figure}

RSMs are also used to extract the lattice parameters of the TF at
room temperature, which are shown in Figure \ref{Lattice-SC-TF} and
compared with those of SC. It can be seen that indeed, in SC, the
effect of Cr and O doping are different. The incorporation of Cr tends
to expand the crystalline lattice showing a discontinuous change in
the lattice parameters. This is an indication of the PM-PI transition
occurring in SC at room temperature when the Cr concentration is varied
from 0.8\% to 1\% \cite{McWhan PRB 1970}. In contrast, for the O-doped
SC, the evolution of the lattice parameters shows a smooth decrease
with increasing oxygen concentration \cite{Ueda JSSC 1980}. Since
we grow on Al\textsubscript{2}O\textsubscript{3} (lattice parameter
\textit{a} = 4.754 Å) an in-plane lattice mismatch will influence
the TF. A mismatch of 4.2\% and up to 5.2\% and 4\% are expected for
the undoped, Cr and O doping cases, respectively. When comparing TF
with Cr-doped SC, it can be noted that they are not entirely relaxed.
The c-axis value for the undoped film is lower than the bulk value
while the a-axis is larger, which arises from the difference in thermal
expansion coefficient between the film and the substrate \cite{Dillemans TSF 2012,Dillemans APL 2014}.
Still the dependence of the c-axis with Cr doping for the TF is in
good agreement with the Cr-doped SC and it can be seen that the values
gradually shift towards the Cr\textsubscript{2}O\textsubscript{3}
bulk value (doping = 1 in the plot). On the other hand, the a-axis
lattice parameter remains substantially smaller due to the larger
mismatch with the substrate for Cr-doped V\textsubscript{2}O\textsubscript{3}.

\begin{figure}
\begin{singlespace}
\includegraphics[scale=0.6]{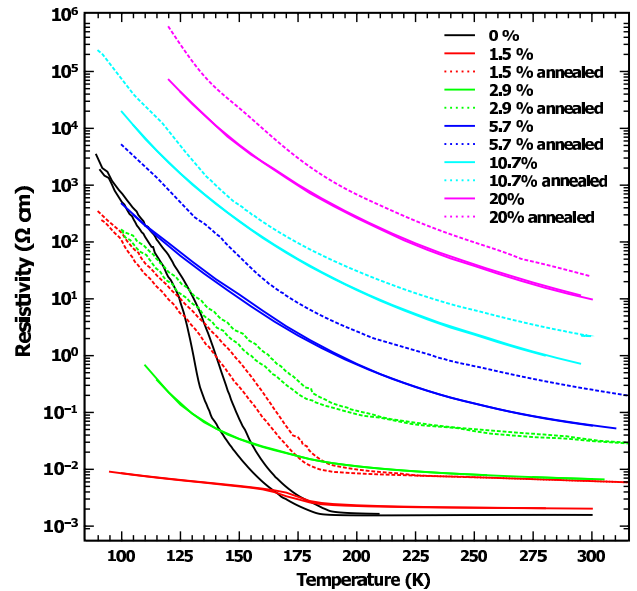}\protect\caption{\label{RvsT-Cr-doped-all}Resistivity versus temperature of Cr-doped
V\protect\textsubscript{2}O\protect\textsubscript{3} TF with different
Cr concentrations. TF as-grown (continuous line) and after vacuum
annealing (dotted line) compared with the undoped V\protect\textsubscript{2}O\protect\textsubscript{3}
case. }
\end{singlespace}
\end{figure}

The resistivity versus temperature in the range from 90 K to 300 K
of the Cr-doped V\textsubscript{2}O\textsubscript{3} TF as-grown
and after the vacuum annealing are shown in Figure \ref{RvsT-Cr-doped-all}.
For the pure V\textsubscript{2}O\textsubscript{3} sample, bulk-like
electrical properties are found as the sharp and well-known MIT at
about 160 K with a resistivity change of 6 decades is clearly observed.
In the as-grown case, the resistivity of the samples with higher Cr
concentrations (> 3\%) increases monotonically when decreasing the
temperature. However, the sample with 2.9\% Cr concentration shows
a much lower resistivity compared with higher Cr concentration alloys,
and strikingly the one with 1.5\% Cr concentration shows a nearly
metallic behavior down to 100 K with the consequent suppression of
the LT insulating state. Compared to the resistivity of the undoped
film at 100 K, this corresponds to a reduction of more than 5 orders
of magnitude. What could be the origin of this drastic reduction? 

Here, we examine three possibilities to explain the collapse of the
LT insulating state. First, it has been reported that hydrogen doping
stabilizes a metallic phase in VO\textsubscript{2} thin films \cite{Wu JACS 2011,Zhao APL 2014}.
However, this type of doping in our Cr-doped samples is very unlikely,
since the H\textsubscript{2} partial pressure measured during the
use of only the Cr cell is one order of magnitude smaller than the
one originated by the V source. The second possibility is the effect
of disorder as recently reported for irradiated V\textsubscript{2}O\textsubscript{3}
TF \cite{Ramirez PRB 2015}. However, in our case, the high temperature
growth process leads to bulk-like (structural and electrical) properties
in pure V\textsubscript{2}O\textsubscript{3} TF suggesting limited
disorder in our films \cite{Dillemans APL 2014}. Finally, the third
option is the addition of oxygen that changes the electrical properties
of V\textsubscript{2}O\textsubscript{3} towards the metallic state
at low temperatures \cite{McWhan JPC 1971,Ueda JSSC 1980,Shivashankar PRB 1983}.
This effect has also been observed in low Cr-doped SC \cite{Kuwamoto-Honig JSSC 1980}.
Hence, we propose that Cr doping promotes oxygen excess in our films
during deposition leading to the collapse of the LT insulating state
at low Cr concentrations.

To confirm this hypothesis, the Cr-doped V\textsubscript{2}O\textsubscript{3}
TF were annealed in vacuum for 5 minutes at the same temperature as
the deposition temperature. We can observe in Figure \ref{RvsT-Cr-doped-all}
that the transport properties of the low Cr-doped samples have changed
significantly upon annealing. The observed large increase in the LTR
-- almost up to the value of the undoped case -- and the presence
of hysteresis, both confirm that the LT insulating state and the MIT
have been recovered for the 1.5\% Cr-doped film. Meanwhile, for the
higher Cr-doped samples after annealing, a shift towards higher resistivities
for the entire curve is observed. 

\begin{figure}
\begin{singlespace}
\includegraphics[scale=0.55]{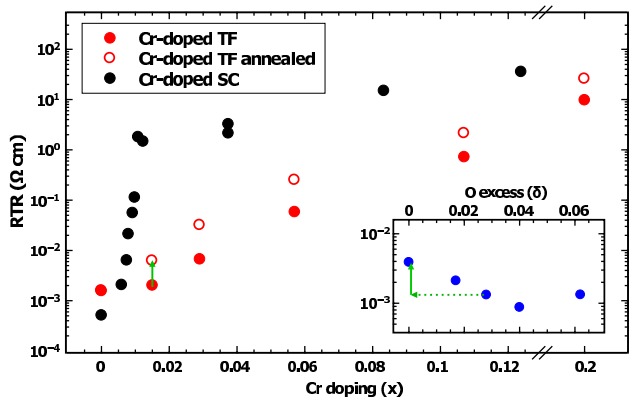}
\end{singlespace}

\protect\caption{\label{RTR}RTR of Cr-doped SC as a function of doping concentration
\cite{McWhan PRB 1970} compared with the as-grown and annealed Cr-doped
TF. Inset shows the RTR versus O excess for V\protect\textsubscript{2}O\protect\textsubscript{3+\textgreek{d}}
SC \cite{Ueda JSSC 1980}. Vertical green arrows indicate the change
in RTR when \textgreek{d} = 0.03 oxygen is removed in SC and the equivalent
change in RTR for the 1.5\% Cr-doped sample after the annealing. }
\end{figure}

In Figure \ref{RTR} the RTR for different Cr doping concentrations
in SC is compared with the corresponding values for the TF in the
as-grown as well as in the annealed state. Again, we observe that
Cr doping is different from O doping in bulk, the RTR increases about
four orders of magnitude with the insertion of only 1\% Cr \cite{McWhan PRB 1970}
while moderate O doping (see inset) has the opposite effect, the RTR
gradually decreases as the system becomes more metallic \cite{Ueda JSSC 1980}.
In our films, the RTR is much more reduced than the bulk case, partially
due to strain that most likely hinders the PM-PI transition to occur.
After annealing in vacuum, we observe an increase of RTR in all the
TF alloys evolving towards the bulk case. It is important to notice
that for the undoped film there is no change in RTR after the annealing
-- the data points in the figure overlap completely -- which is consistent
with stoichiometric V\textsubscript{2}O\textsubscript{3} films.
Furthermore, for the lowest Cr concentration, the change in RTR is
about 3$\times$10\textsuperscript{-3} \textgreek{W} cm, which is
nearly the same as the one upon removal of 0.03 oxygen excess in non-stoichiometric
SC \cite{Ueda JSSC 1980}. These changes are indicated with the vertical
green arrows in the figure. Hence, we can estimate that the amount
of oxygen lost during the annealing -- for the lowest Cr-doped sample
-- corresponds to about 1\%. 

When Cr doping is added, the 24\% increment (from 4.2 to 5.2\%) in
the in-plane lattice mismatch with the substrate needs also to be
accommodated by the TF. The insertion of oxygen would thus lead to
a smaller mismatch and can explain its preferential incorporation
when Cr-doped TF are grown. There are two other mechanisms which can
contribute to an increase of the oxygen content in Cr-doped films:
(1) the larger electron affinity of Cr atoms in comparison with V
atoms; (2) the increased average lattice spacing of Cr-doped films
allows more oxygen atoms to be incorporated during growth. This additional
oxygen (or cation vacancies) is later removed by the vacuum annealing
at high temperature. 

However, after the annealing, the properties both structurally as
well as electrically do not yet reach those of the equivalent bulk
compounds. In SC, the change in lattice parameter (see Figure \ref{Lattice-SC-TF})
across the PM-PI transition can account at least partially for the
resistivity jump across the transition. The \textit{\textcolor{black}{a}}
lattice parameter as well as the lattice volume = \textit{\textcolor{black}{a}}\textsuperscript{2}\textit{\textcolor{black}{c}}
\textit{\textcolor{black}{sin}}(60\textsuperscript{o}) both increase
by about 1\% , which leads to a reduction of the orbital overlap and
thus a decrease of bandwidth (or, equivalently, a strengthening of
the electron correlations). Note that the change in lattice parameters
in our films after the annealing is very small (less than 0.05\%,
not shown here). Therefore, the absence of this transition even after
the vacuum annealing may be attributed to the large in-plane mismatch
and the clamping to the substrate, which prevents the in-plane lattice
parameter to change in the TF as it is observed in bulk \cite{McWhan PRB 1970}.
In a previous work \cite{Dillemans APL 2014} it has been shown that
the MIT in V\textsubscript{2}O\textsubscript{3} layers grown directly
on Al\textsubscript{2}O\textsubscript{3 }is lost for thicknesses
below 5 nm due to the presence of a large strain; however, the transition
can be recovered when a thin Cr\textsubscript{2}O\textsubscript{3 }buffer
layer is inserted. Then, the growth of TF alloys on substrates with
a larger in-plane lattice parameter may help to further understand
the present results. 

In conclusion, we demonstrate that high quality Cr-doped V\textsubscript{2}O\textsubscript{3}
TF can be grown epitaxially. We report structural and electrical properties
of TF alloys with Cr concentrations up to 20\% grown by MBE. For the
highly doped samples (> 3\%), an increase of the resistance with decreasing
temperature is measured. Strikingly, for the low doped samples (<3\%),
a collapse of the insulating state is observed with a reduction of
the low temperature resistivity by up to 5 orders of magnitude. Using
a vacuum annealing procedure at high temperature, the LT insulating
state is recovered for these films showing evidence that an oxygen
excess introduced during the film growth is responsible for the observed
collapse. Therefore, our results also demonstrate that co-doping with
Cr as well as with O is possible in TF. This gives to the V\textsubscript{2}O\textsubscript{3}
system \emph{two handles working in opposite directions}. Furthermore,
since oxygen can be mobile under the application of an electric field
\cite{Jeong Sci 2013,Quintero PRL 2007,Yoshida APL 2008}, these results
suggest that the Cr-O-V system holds a great potential for different
electronic devices. Moreover, it will be interesting to investigate
the resistive switching behavior in our films in vertical Cr-doped
V\textsubscript{2}O\textsubscript{3} based structures and in different
gas environments as well.

The authors acknowledge financial support from the FWO Project No.
G052010N10 as well as the EU-FP7 SITOGA Project. P.H. acknowledges
support from Becas Chile - CONICYT.

\end{document}